\documentclass[prd,tightenlines,superscriptaddress]{revtex4}
\usepackage{enumerate}
\usepackage{amssymb}
\usepackage{amsfonts}
\usepackage{mathrsfs}
\usepackage{latexsym}
\usepackage{bm}
\usepackage{amscd}
\usepackage{graphicx}
\usepackage{epsfig}
\usepackage{hhline,multirow}
\usepackage{dcolumn}
\usepackage{url}
\usepackage[colorlinks=true,linkcolor=blue]{hyperref}
\usepackage{color}
\usepackage{indentfirst}
\usepackage{subfigure}
\usepackage{psfrag}
\usepackage{slashed}
\usepackage{float}
\usepackage{setspace}

\begin{document}

\title{Pauli form factors of electron and muon in nonlocal quantum electrodynamics }

\author{Fangcheng He}
\affiliation{Institute of High Energy Physics, CAS,
                Beijing 100049, China}

\author{P. Wang}
\affiliation{Institute of High Energy Physics, CAS,
                Beijing 100049, China}
\affiliation{Theoretical Physics Center for Science Facilities, CAS,
                Beijing 100049, China}

\date{\today}

\begin{abstract}

Pauli form factors of electron and muon are studied in nonlocal quantum electrodynamics.
We calculate one loop QED correction to their Pauli form factors. The relativistic regulator 
is generated by the correlation function in the nonlocal interaction. The cut-off parameter 
$\Lambda$ in the regulator is determined to get the consistent anomalous magnetic moments
of electron and muon at the same level as local QED. When momentum transfer is large, 
there exists obvious difference between nonlocal and local QED. 

\end{abstract}


\maketitle

\section{Introduction}

The anomalous magnetic moments of electron and muon $a_e$ and $a_\mu$ are among the 
most precisely determined observables in particle physics. The most accurate measurement 
of $a_e$ so far has been carried out by the Harvard group as 
$a_e=1159652180.73(28)\times 10^{-12}$ \cite{Hanneke:2008tm,Hanneke:2010au}.
Further improvements for the electron and positron measurements are currently prepared by the 
Harvard group \cite{Hoogerheide:2014mna}.
For the muon magnetic moment, the E821 measurement at BNL \cite{Bennett:2006fi}, corrected for updated 
constants \cite{Mohr:2008fa,Grange:2015fou}, is $a_\mu=116592089(63) \times 10^{-11}$.
Two next generation muon $g-2$ experiments at Fermilab in the US
and at J-PARC in Japan have been designed to reach a four times better precision from 0.54 ppm 
to 0.14 ppm \cite{Holzbauer:2017ntd}.

The standard model (SM) values of lepton anomalous magnetic moments are estimated separately by theorists. 
It has several parts: quantum electrodynamics (QED), electroweak (EW), 
hadronic vacuum polarization (HVP) and hadronic light-by-light (HLbL). 
The QED and EW parts are known very well perturbatively. 
For example, the QED contribution to the anomalous magnetic moment of electron and 
muon is known up to 5-loop order \cite{Tanabashi:2018oca}.
The others, being hadronic terms, are less well known and are estimated using
various techniques, including data from other experiments \cite{Keshavarzi:2018mgv} 
and lattice calculations \cite{Blum:2018mom,Chakraborty:2018iyb}. 

With the updated theoretical prediction from standard model and the existing experimental 
measurement, one can get the discrepancy. For example, for electron, the discrepancy 
is \cite{Aoyama:2017uqe,Davoudiasl:2018fbb}
\begin{equation}
\Delta a_e = (-87 \pm 36) \times 10^{-14},
\end{equation}
which shows a $2.4 \sigma$ deviation.
For muon, the anomalous magnetic moment $a_\mu$ has 3.7 $\sigma$ discrepancy with a positive sign, 
opposite to the $a_e$ deviation \cite{Blum:2018mom,Jegerlehner:2018zrj}. The value of $\Delta a_\mu$ is
\begin{equation}
\Delta a_\mu = (2.74 \pm 0.73) \times 10^{-9}.
\end{equation}
As standard model predictions almost without exception match perfectly all other experimental 
information, the deviation in one of the most precisely measured 
quantities in particle physics remains a mystery and inspires the
imagination of model builders. In this paper, we will investigate the anomalous form factors
of electron and muon with the nonlocal quantum electrodynamics inspired from the nonlocal 
effective field theory (EFT). 

Nonlocal effective field theory was recently proposed to study hadron physics. Different from
the traditional EFT, the interaction is nonlocal which reflects the non-point behavior
of hadrons. Therefore, in the nonlocal Lagrangian, there is a correlation function $F(x-y)$,
where the baryon is located at $x$ and meson at $y$. If $F(x-y)$ is chosen to be $\delta(x-y)$, 
the Lagrangian will be changed back to the local one. With the correlation function, there is no
ultraviolet divergence in the loop intergral. The nonlocal EFT has been applied to study the nucleon 
electromagnetic form factors, strange form factors, parton distributions, etc \cite{He:2017jtk,He:2017viu,He:2018eyz,Salamu:2018cny}. 

The nonlocal Lagrangian may be not only the phenomenological method to deal with the divergence,
but also the general property for all the physical interactions. In other words, whether there
is divergence or not, the nonlocal regulator always exists. Therefore,
with the same idea, the interaction between electron and photon could also be nonlocal.
In the classical scenario (tree level diagram), the nonlocal effect is certainly negligible for 
the low momentum transfer. However, for the quantum fluctuation or loop diagram, the
internal photon can detect the structure of the physical particle since its momentum can be
infinite. 

In this paper, we will show the deviation of the lepton Pauli form factors 
in nonlocal QED from local one. It can be seen that high order QED corrections and hadronic effects
do not affect the final conclusion since the deviation is significant, especially at 
large $Q^2$. Compared with the observables in hadron physics, where the non-perturbative effect
is very important, it is a great advantage for the lepton anomalous form factors to serve as 
the quantities to test standard model. In the following, we will first derive the form factors in nonlocal QED
and then show the numerical results. 

\begin{figure}[t]
\begin{center}
\includegraphics[scale=0.75]{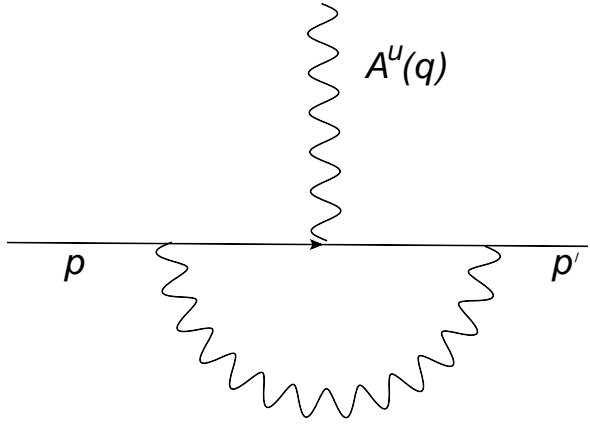}
\caption{Feynman diagram for the one loop vertex correction to the lepton form factors.}
\label{diagrams}
\end{center}
\end{figure}

\section{Nonlocal QED Lagrangian}
The nonlocal Lagrangian for quantum electrodynamics is written as
\begin{equation}
\mathcal{L}_{QED}^{nl} = \bar{\psi}(x)(i\slashed{\partial}-m)\psi(x)-\frac14F_{\mu\nu}(x)^2
-e\int\!\,d^4 a\bar\psi(x)\slashed{A}_\mu(x+a)\psi(x)F(a),
\end{equation}
where the electron field $\psi(x)$ is located at $x$ and the photon field $A_\mu(x+a)$ is located at
$x+a$. $F(a)$ is the correlation function normalized as $\int d^4 a F(a)=1$. If it is chosen to be a $\delta$ function, 
the nonlocal Lagrangian will be changed back to the local one. The above nonlocal Lagrangian is invariant
under the following gauge transformation
\begin{equation}
\psi(x)\rightarrow\,e^{i\alpha(x)}\psi(x),~~~~~~~~A_\mu\rightarrow\,A_\mu-\frac1e\partial_\mu\,\alpha^\prime(x),
\end{equation}
where $\alpha(x)=\int\,da\alpha^\prime(x+a)F(a)$. 
Different from the nonlocal Lagrangian for EFT, where the gauge link is introduced to guarantee the local
gauge invariance, here we need no gauge link since photon is charge neutral.

With the correlation function, the lepton-photon interaction is momentum dependent as $e\gamma_\mu\,\tilde F(q)$,
where $\tilde F(q)$ is Fourier transformation of the correlation function $F(a)$ and 
$q$ is photon momentum. Ward-Takahashi identity becomes
\begin{equation}\label{eq:A}
-iq_\mu\Gamma^\mu(p+q,p)=\tilde F(q)(S^{-1}(p+q)-S^{-1}(p)),
\end{equation}
where  $\Gamma^\mu(p+q,p)$ is the vertex.
$S(p+q)$ and $S(p)$ are the lepton propagators with wave function renormalization
\begin{equation}
S(p+q)=\frac{iZ_2}{\slashed{p}+\slashed{q}-m},~~~~~~~~~S(p)=\frac{iZ_2}{\slashed p-m},
\end{equation}
where $Z_2$ is lepton wave function renormalization factor. 
At $q=0$, $\tilde F(q)$ is 1 due to the normalization of $F(a)$ and Eq.(\ref{eq:A}) can be written as 
\begin{equation}\label{eq:r0}
Z_2\Gamma^\mu(p,p)=\gamma^\mu \tilde {F}(0).
\end{equation}
The lepton form factors is defined as \cite{Peskin:1995ev}
\begin{equation}\label{eq:rn}
Z_2\Gamma^\mu(p+q,p)=\gamma^\mu F_1(q^2)+\frac{i\sigma^{\mu\nu}q_\nu}{2m}F_2(q^2).
\end{equation}
Therefore, we have $F_1(0)=\tilde {F}(0)$ which is consistent with that the renormalized lepton
charge is 1.

The one loop Feynman diagram for the lepton form factors is plotted in Fig.~1. At one loop level,
the vertex is written as
\begin{eqnarray}\label{eq:loop}
\bar{u}(p')\Gamma^\mu_{loop}(p',p) u(p)=\bar{u}(p')\int\!\frac{d^4k}{(2\pi)^4}\tilde F(q) \tilde F(k)^2(-ie\gamma^{\nu})\frac{i}{\slashed{p'}-\slashed{k}-m}\gamma^\mu\frac{i}{\slashed{p}-\slashed{k}-m}(-ie\gamma^{\rho})\frac{-ig_{\nu\rho}}{k^2}u(p)
\end{eqnarray}
From Eq.~(\ref{eq:rn}), one can get
\begin{eqnarray}
F_1^{loop}(q^2)&=&\frac{-ie^2\tilde F(q^2)}{(4m^2-q^2)^2}\int\!\frac{d^4 k}{(2\pi)^4}\tilde F^2(k^2) \big\{\frac{-24m^2((k\cdot p)^2+(k\cdot p')^2)+8m^2k^2(4m^2-q^2)}{((p'-k)^2-m^2)(p-k)^2-m^2)k^2}\nonumber    \\ 
&+&\frac{2(2m^2-q^2)(4m^2-q^2)^2-4(4m^2+2q^2)(k\cdot p)(k\cdot p')-4(k\cdot p+k\cdot p')(8m^4-6m^2q^2+q^4)}{((p'-k)^2-m^2)(p-k)^2-m^2)k^2}\big\}
\end{eqnarray}
and
\begin{eqnarray}
F_2^{loop}(q^2)&=&\frac{-ie^2\tilde F(q^2)8m^2}{q^2(4m^2-q^2)^2}\int\!\frac{d^4k}{(2\pi)^4}F^2(k^2)\big\{ \frac{(4m^2+2q^2)((k\cdot p)^2+(k\cdot p')^2)-8(m^2-q^2)(k\cdot p)(k\cdot p')}{((p'-k)^2-m^2)(p-k)^2-m^2)k^2} \nonumber    \\ 
&+&\frac{(q^4-4m^2q^2)(k\cdot p+k\cdot p'+k^2)}{((p'-k)^2-m^2)(p-k)^2-m^2)k^2}\big\}.
\end{eqnarray}
In the above expressions, the momentum dependent vertexes $\tilde F(q)$ and $\tilde F(k)$ appear.
For $\tilde F(q)$, if the external momentum $q$ is much smaller than the scale of electron, the size of
electron can be neglected and $\tilde F(q) \simeq 1$. However, for $\tilde F(k)$, the internal momentum
$k$ varies from 0 to infinity. The regulator is very important and it makes the loop integral for $F_1$ and $F_2$ 
both convergent.

For the Dirac form factor $F_1(q^2)$, there is a contribution from tree level as $ Z_2 \tilde {F} (q^2)$.
The wave function renormalization constant $Z_2$ is obtained as \cite{Peskin:1995ev}
\begin{eqnarray}
Z_2-1&=&\frac{d\Sigma(\slashed{p})}{d\slashed{p}}\bigg |_{\slashed{p}=m},
\end{eqnarray}
where the lepton self-energy is expressed as 
\begin{eqnarray}
\Sigma(\slashed{p})=-\int\!\frac{d^4k}{(2\pi)^4} \tilde F^2(k)(ie\gamma^{\nu})\frac{1}{\slashed{p}-\slashed{k}-m}(-ie\gamma^{\rho})\frac{-ig_{\nu\rho}}{k^2}.
\end{eqnarray}
It is straightforward to find $\frac{d\Sigma(\slashed{p})}{d\slashed{p}} = -F_1^{loop} (0)$.
Again, we have $F_1(0) = Z_2 +F_1^{loop}(0) = 1$.

In the following, we focus on the Pauli form factor $F_2(q^2)$ and $F_2(0)$ gives the anomalous magnetic moment.
For the numerical calculation, the regulator $\tilde F (k^2)$ is chosen to be a dipole form as
\begin{equation}
\tilde F(k^2)=\frac{\Lambda^4}{(k^2-\Lambda^2)^2}.
\end{equation} 
The Pauli form factor at one loop level can be obtained as
\begin{eqnarray}\label{eq:nonlocal}
F_2^{nl}(Q^2)&=&\frac{\alpha}{2\pi}\tilde F(-Q^2)\int_0^1\,dx\int_0^{1-x}dy\frac{2\Lambda^{8}m^2(x+y)(1-x-y)^5}{(Q^2xy+m^2(x+y)^2)(m^2(x+y)^2+Q^2xy+(1-x-y)\Lambda^2)^4}.
\end{eqnarray}
When $\Lambda$ goes to infinity, the local result will be recovered as
\begin{eqnarray}\label{eq:local}
F_2^{lo}(Q^2)&=&\frac{\alpha}{2\pi}\int_0^1\,dx\int_0^{1-x}dy\frac{2m^2(1-x-y)(x+y)}{m^2(x+y)^2+Q^2xy},
\end{eqnarray}
which results in the well known anomalous magnetic moment at one loop level $a_l=F_2^{lo}(0)=\frac{\alpha}{2\pi}$
\cite{Schwinger:1948iu}. 
In the above equations, we replaced $q^2$ by $-Q^2$ for convenience.
With the results from nonlocal and local QED, we can get the discrepancy between them as
$\Delta F_2= F_2^{lo} - F_2^{nl}$. The relative discrepancy is defined as $R =\Delta F_2 / F_2^{lo}$.
The loop integral for Pauli form factors are both ultraviolet convergent in local and nonlocal cases.
We should mention that the regulator is not introduced phenomenologically to deal with the divergence.
This is different from the original finite-range-regularization \cite{Young:2002ib,Leinweber:2003dg,
Wang:2007iw,Wang:2010hp,Allton:2005fb,Armour:2008ke,Hall:2013oga,Leinweber:2004tc,Wang:1900ta,Wang:2012hj,
Wang:2013cfp,Hall:2013dva,Wang:2015sdp,Li:2015exr,Li:2016ico,Wang:2008vb}. 
The regulator is naturally generated from the nonlocal Lagrangian with the naive idea that the interaction between
photon and lepton does not necessary take place at one point. For the ultraviolet divergent integral at local case,
the regulator will make the integral convergent. For the integral which is convergent at local case, 
the regulator also exists and will give obvious deviations from the local result, especially at large momentum transfer.

\begin{figure}[tbp]
\begin{center}
\includegraphics[scale=0.85]{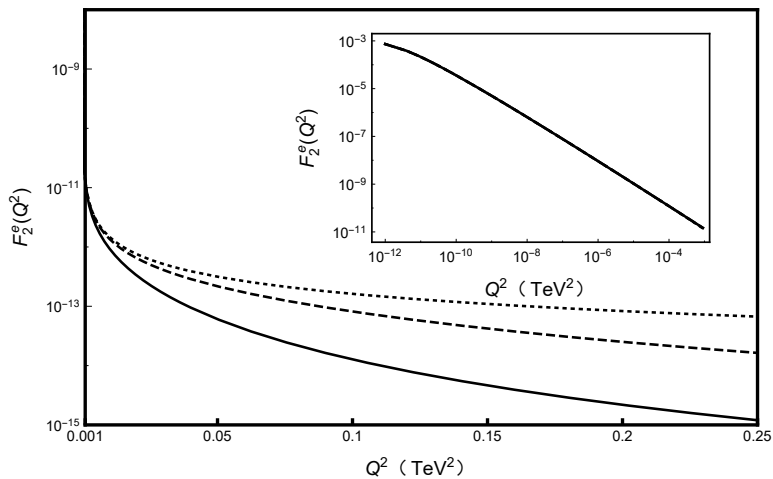}
\caption{Electron Pauli form factor $F_2^e (Q^2)$ versus momentum transfer $Q^2$. The solid, dashed and dotted lines are for $\Lambda=0.2$, $0.5$ TeV and local limit, respectively.
The small figure is for $F_2^2(Q^2)$ at low $Q^2$ up to $0.001$ TeV$^2$.}
\label{diagrams}
\end{center}
\end{figure}

\section{Numerical Results}
In the numerical calculation, there is one free parameter $\Lambda$ in the regulator needs to be determined. 
$\Lambda$ is order of 1 GeV for nucleon. For leptons, $\Lambda$ could be much larger
since their sizes are much smaller. Certainly, on the one hand, the smaller the value of $\Lambda$, 
the larger the deviation from the standard model. On the other hand, $\Lambda$ should be
large enough and make the nonlocal results consistent with the experiments at the same level
as standard model. When $\Lambda=0.2$ TeV, the calculated $a_e^{nl}$ in nonlocal QED is 0.00116171491307, 
which is $2.0 \times 10^{-14}$ deviation from the corresponding value in standard model. 
Considering the experimental accuracy $2.6 \times 10^{-13}$ \cite{Tanabashi:2018oca} and discrepancy between
experiments and SM prediction $8.7 \times 10^{-13}$, the choice of $\Lambda=0.2$ TeV
is fine. For muon, $a_\mu$ in nonlocal QED is $8.6 \times 10^{-10}$
deviation from that in local case. Comparing with $\Delta a_\mu$ is $2.7 \times 10^{-9}$,
$\Lambda = 0.2$ TeV is also fine for muon. Therefore, in the numerical calculation, we
show the results with $\Lambda =0.2$ and $0.5$ TeV. In principle, 
QED itself can not determine the form of the regulator and the value of $\Lambda$. It can only be 
determined by the experimental data, especially at finite $Q^2$. We will show that whatever $\Lambda$ is, 
the relative discrepancy between local and nonlocal QED is always significant if momentum transfer $Q^2$ 
is large enough.

In Fig. 2 we plot the Pauli form factor of electron $F_2^e(Q^2)$ versus $Q^2$.
The solid, dashed and dotted lines are for $\Lambda=0.2$, $0.5$ TeV and local limit, respectively.
The small figure inside Fig.~1 is to show the form factor at low $Q^2$.
It can be seen that the form factor decreases very fast with the increasing $Q^2$ due to
the fact that the electron mass is very small. When $Q^2$ is small, the discrepancy between
nonlocal QED and SM is much smaller than the form factors themselves. For example,
at $Q^2=0$, the discrepancy $\Delta F_2^e$ is at least $10^{11}$ times smaller than the anomalous magnetic 
moments. With the increasing $Q^2$, the discrepancy is clearly shown in the figure
when its value is comparable with the form factors. Due to the nonlocal effect, the form factor 
in nonlocal QED is smaller than that in SM at any $Q^2$.

\begin{figure}[t]
\begin{center}
\includegraphics[scale=0.85]{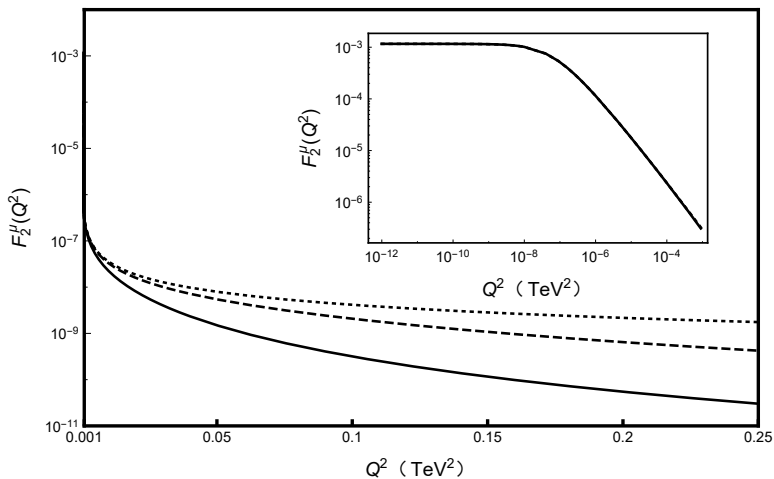}
\caption{Same as Fig.2 but for muon Pauli form factor.}
\label{diagrams}
\end{center}
\end{figure}

\begin{figure}[t]
\begin{center}
\includegraphics[scale=0.85]{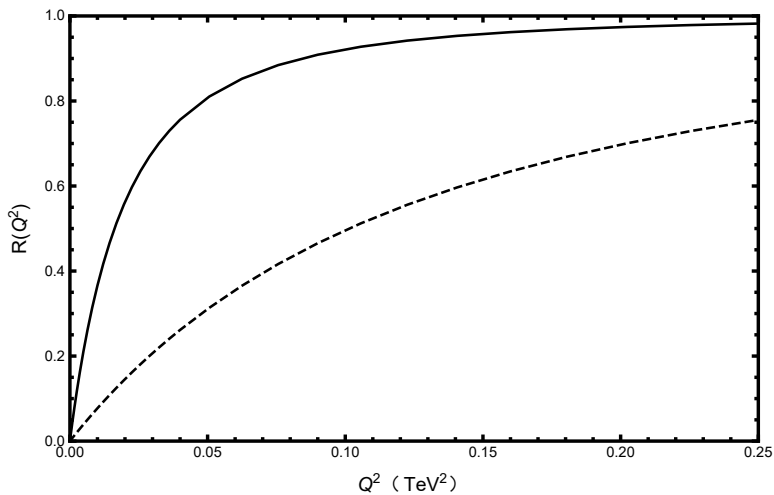}
\caption{The relative deviation $R(Q^2)$ for electron and muon versus $Q^2$. The solid and dashed lines are 
for $\Lambda = 0.2$ and $0.5$ TeV, respectively.}\label{diagrams}
\end{center}
\end{figure}

The result for muon is shown in Fig.~3. Similarly as for electron, the form factor drops
quickly with the increasing $Q^2$. Since the mass of muon is larger than that of electron,
the form factor drops slower than electron. At $Q^2=0$, the form factors of electron and
muon are close to each other (In SM, they are exactly the same at leading order). However, 
at finite $Q^2$, they have huge difference. For example, for $0.001 < Q^2 < 0.01$ TeV$^2$, 
the muon form factor is about $4-5$ magnitude larger than electron form factor. The discrepancy 
between nonlocal QED and SM is also $4-5$ magnitude larger for muon than that for electron.

To see clearly the discrepancy between nonlocal QED and SM, we plotted the relative deviation
$R$ in Fig.~4. The solid and dashed lines are for $\Lambda = 0.2$ and $0.5$ TeV, respectively.
The discrepancy $\Delta F_2$ and the form factor $F_2$ are both much larger for muon than that
for electron. The relative deviation $R$ are almost the same for electron and muon.
This can be seen from Eqs.~(\ref{eq:nonlocal}) and (\ref{eq:local}). The lepton mass dependence in the
numerator cancels each other for $F_2^{nl}/F_2^{lo}$. The mass in the denominator
is much smaller than $\Lambda$ which makes $R$ nonsensitive to the mass.
At $Q^2=0$, the relative deviation is very small and is order of $10^{-11}$ and $10^{-7}$ for
electron and muon. This is more or less the same order for the relative deviation of SM from the experiments.
Therefore, to confirm this discrepancy for electron, the experiment should be very accurate to get
10 more effective digits. The relative deviation $R$ increases with the increasing $Q^2$.
For example, at $Q^2=0.01$ TeV$^2$, $R$ is about 0.37 and 0.08 for $\Lambda=0.2$ and 0.5 TeV,
respectively. When $Q^2$ is larger than 0.1 TeV$^2$, $R$ is larger than 0.5 for both $\Lambda$s.
This means the nonlocal value of $F_2^{nl}$ is less than half of the SM value. For even larger $Q^2$,
say larger than 0.2 TeV$^2$, $F_2^{nl}$ could be one magnitude smaller than the SM value.
We can see though the absolute value of the form factor is small, the relative deviation of nonlocal QED from SM
is very large. Even if the experiment can only measure the form factor with one effective digit at finite $Q^2$,
one can still conclude whether there is physics beyond SM. Since the deviation is 
so large at finite $Q^2$, the conclusion is not changed by the high order QED correction or hadronic effect,
as we know for both electron and muon,
these corrections are less than one percent. In summary, we list the discrepancy between nonlocal QED and SM 
and the relative deviation at some $Q^2$ for electron and muon in Table I. 

\begin{table}
\caption{The discrepancy between nonlocal QED and SM and the relative deviation for electron and muon.}
\begin{center}
\begin{ruledtabular}
\begin{tabular}{c|cccccc}
$Q^2$ (TeV$^2$) & 0 & $0.001$ & $0.01$ & $0.05$ & $0.1$ & $0.2$ \\ 
\hline 
$\Delta F_2^e$ ($\Lambda=0.2$ TeV) & $ 2.03\times 10^{-14}$ & $ 6.63\times 10^{-13}$ & $ 5.41\times 10^{-13}$ &
$ 2.55\times 10^{-13}$ & $ 1.49\times 10^{-13}$ & $ 8.08\times 10^{-14}$ \\ 
$\Delta F_2^e$ ($\Lambda=0.5$ TeV) & $ 3.27\times 10^{-15}$ & $ 1.10\times 10^{-13}$ & $ 1.14\times 10^{-13}$ & 
$ 9.79\times 10^{-14}$ & $ 8.03\times 10^{-14}$ & $ 5.79\times 10^{-14}$ \\ 
$R^e$ ($\Lambda=0.2$ TeV) & $1.74 \times 10^{-11}$ & 0.050 & 0.370 & 0.808 & 0.922 & 0.973 \\ 
$R^e$ ($\Lambda=0.5$ TeV) & $2.81 \times 10^{-12}$ & 0.008 & 0.077 & 0.311 & 0.496 & 0.697 \\ 
\hline 
$\Delta F_2^\mu$ ($\Lambda=0.2$ TeV) & $ 8.65\times 10^{-10}$ & $ 1.50\times 10^{-8}$ & $ 1.32\times 10^{-8}$ & 
$ 6.45\times 10^{-9}$ & $ 3.84\times 10^{-9}$ & $ 2.11\times 10^{-9}$ \\ 
$\Delta F_2^\mu$ ($\Lambda=0.5$ TeV) & $ 1.38\times 10^{-10}$ & $ 2.49\times 10^{-9}$ & $ 2.80\times 10^{-9}$ & 
$ 2.50\times 10^{-9}$ & $ 2.08\times 10^{-9}$ & $ 1.52\times 10^{-9}$ \\ 
$R^\mu$ ($\Lambda=0.2$ TeV) & $7.44 \times 10^{-7}$ & 0.051 & 0.371 & 0.812 & 0.924 & 0.975 \\ 
$R^\mu$ ($\Lambda=0.5$ TeV) & $1.19 \times 10^{-7}$ & 0.008 & 0.079 & 0.314 & 0.500 & 0.701 \\ 
\end{tabular}
\end{ruledtabular}
\end{center}
\end{table}

\section{Summary}

We studied the anomalous form factors of electron and muon in nonlocal QED inspired by the nonlocal effective 
field theory for hadron physics. The interaction between lepton and photon is describe by the nonlocal QED.
The regulator is generated from the correlation function in the nonlocal Lagrangian. For the Dirac form factor 
of leptons and electromagnetic form factors of nucleon, the ultraviolet divergence of loop integral in local interaction will 
disappear with the regulator. For the Pauli form factors of electron and muon, 
the loop integrals are both convergent for nonlocal and local QED. The parameter $\Lambda$ is
chosen to make the nonlocal result of lepton anomalous magnetic moments consistent with the experimental 
data at the same level as SM.
The form factors of the electron and muon decrease very fast with the increasing $Q^2$ because of
the small lepton masses. At $Q^2=0$, the absolute discrepancy between nonlocal and local results
is more or less similar as the discrepancy between experiments and SM predictions.
However, the relative deviation from the SM is large at finite $Q^2$. Since the high order 
QED correction and hadronic effect is less than one percent, the large relative deviation
is not affected. If this deviation can be measured, it will definitely indicate new physics beyond SM.
It will also imply that nonlocal behavior could be the general property for all the interactions and 
as a result, we will have to reconsider the regularization and renormalization.

\section*{Acknowledgments}

This work is supported by the National Natural Sciences Foundations of China under the grant No. 11475186,
the Sino-German CRC 110 ``Symmetries and the Emergence of Structure in QCD" project by NSFC under the grant
No.11621131001, and the Key Research Program of Frontier Sciences, CAS under grant No. Y7292610K1.


\providecommand{\href}[2]{#2}\begingroup\raggedright\endgroup

\end{document}